\documentclass[ reprint, amsmath,amssymb, aps,pre,showpacs]{revtex4-1}

\usepackage{graphicx}
\usepackage{dcolumn}
\usepackage{bm}
\usepackage{natbib}
\usepackage{amsmath}
\usepackage{hyperref}

\usepackage[dvipsnames]{xcolor}

\begin{document}

\preprint{APS/123-QED}

\title{Cascading failures in scale-free interdependent networks}

\author{Malgorzata Turalska}
\affiliation{Network Science Division, Army Research Laboratory, Adelphi, MD, USA, 20783}

\author{Keith Burghardt}
\affiliation{Information Sciences Institute, University of Southern California, Marina del Rey, California, USA, 90292}

\author{Martin Rohden}
\affiliation{Department of Computer Science, University of California, Davis, California, USA, 95616}

\author{Ananthram Swami}
\affiliation{Computational and Information Science Directorate, Army Research Laboratory, Adelphi, MD, USA, 20783}

\author{Raissa M. D'Souza}
\affiliation{Department of Computer Science, University of California, Davis, California, USA, 95616}
\affiliation{Department of Mechanical and Aerospace Engineering, University of California, Davis, California, USA, 95616}
\affiliation{Santa Fe Institute, Santa Fe, New Mexico, USA, 87501}

\date{\today}

\begin{abstract}
Large cascades are a common occurrence in many natural and engineered complex systems. In this paper we explore the propagation of cascades across networks using realistic network topologies, such as heterogeneous degree distributions, as well as intra- and interlayer degree correlations. We find that three properties, scale-free degree distribution, internal network assortativity, and cross-network hub-to-hub connections, are all necessary components to significantly reduce the size of large cascades in the Bak-Tang-Wiesenfeld sandpile model. We demonstrate that correlations present in the structure of the multilayer network influence the dynamical cascading process and can prevent failures from spreading across connected layers. These findings highlight the importance of internal and cross-network topology in optimizing stability and robustness of interconnected systems.

\end{abstract}

\maketitle

\section{Introduction}

Occasionally, natural as well as man-made complex networks suffer massive cascades, which are initialized by a breakdown of a small portion of the entire system. Such cascades characterize a plethora of complex phenomena, including neural avalanches \cite{Levina2007,Goh2003,Beggs2003,Chialvo2010}, blackouts in power grids \cite{Carreras2002,Weng2006,Hines2009}, secondary extinctions in ecological systems \cite{Estes2011,Motter2011}, and systemic default of financial institutions \cite{Battiston2012,Bardoscia2017}. Recently, research has focused on how modular structures or interconnections between networks affect large cascades for simple regular network topologies\cite{Brummitt2012}. More realistic topological features, such as broad-scale degree distributions \cite{Eguiluz2005,Bullmore2009}, assortativity \cite{Bialonski2013,Geier2015}, or non-random inter-connectivity between communities \cite{Heuvel2011,Reis2014,Dosenbach2007,Betzel2017}, have yet to be explored. These structural features are seen, for example, in functional brain networks.

Here we develop a systematic study to fill this gap. Our goal is to identify near-optimal architectures for preventing cascading failures in realistic interconnected systems. We demonstrate that  as well as interlayer degree correlations play crucial roles in affecting the occurrence of catastrophic cascades in interdependent heterogeneous networks. In particular we show that vulnerability of individual nodes to fail correlates with degree of assortativity present in the network. This behavior illustrates the importance of considering higher-order network properties when maximizing robustness of interconnected systems.

We study failure cascades with the Bak-Tang-Wiesenfeld (BTW) sandpile model, which self-organizes to an apparent critical state, in which cascades sizes are distributed as a power law \cite{Bak1987,Bak1988,Bonabeau1995,Goh2003,Hoore2013,cajueiro2010,Brummitt2012,Noel2013,Noel2014}. This characteristic mimics the heavy-tailed distributions of failure cascades seen in electrical blackouts \cite{Carreras2002,dobson2007,Weng2006}, neuronal avalanches \cite{Beggs2003}, earthquakes \cite{Geller1997,Sornette2004} and forest fires \cite{christensen1993,sinha2000}. Furthermore, the BTW model captures a common feature of many systems in which individual elements carry a load, but have a fixed capacity \cite{Brummitt2012}. This property makes the BTW model a valuable tool when studying how network cascades result from the individual elements failing due to exceeding their capacity and shedding their load to neighboring elements.

The universality of the cascade size statistics observed in numerous dynamical systems poses significant challenges to the design of strategies to control the occurrence of large catastrophic failures. Namely, because independent of whether the underlying network topology is homogeneous or heterogeneous, failure sizes are characterized by heavy-tailed distributions, thus reducing cascade sizes is not easily achieved through changing connectivity alone \cite{Motter2017}. Brummitt \textit{et al.} \cite{Brummitt2012}, however, demonstrated that  connections between networks act as a control mechanism regulating frequency of catastrophic failures in coupled random regular networks. Here one network can minimize the likelihood of a large cascade by forming an intermediate amount of connections with the other network. Changing the connectivity away from this point enhances the likelihood of large cascades.

Our work extends upon Brummitt \textit{et al.} by coupling together scale-free networks, rather than random regular networks, to better approximate many natural and man-made systems \cite{Moreno2014}. We show that the probability of large cascades is significantly affected by the degree distribution, the inter-layer degree correlations, and the intra-layer degree correlations. These results could not have been obtained without exploring the dynamics of the BTW model on heterogeneous coupled networks, and may provide insight into the evolutionary advantages of particular network structures seen in nature.

We organize the rest of the paper as follows. We begin in Sec.~\ref{sec:sandpile_btw} with a brief background on the sandpile process on individual complex networks. In Sec.~\ref{sec:one_over_k} we show fundamental disparities between results of numerical simulations and assumptions made by branching process approximations to the sandpile model, which motivates discussing the BTW model for large cascades. In Sec.~\ref{sec:sandpile_conn} we study the spread of large cascades through interconnected networks. Finally, we discuss our findings in Sec.~\ref{sec:discussion}.

\section{Sandpile process on isolated complex networks} \label{sec:sandpile_btw}

\begin{figure}[t]
\includegraphics[width=8.6cm]{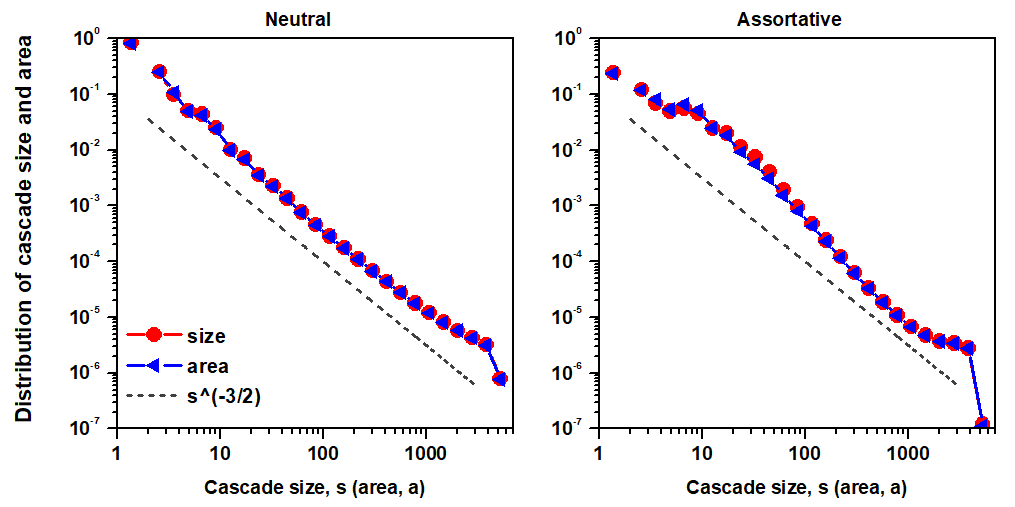}
\caption{Probability distribution of cascade size, $P(s)$ (red circles), and cascade area, $P(a)$ (blue triangles), for a neutral (left panel) and assortative (right panel) scale-free network. The dashed line corresponds to the mean-field solution to the BTW model, where $P(s)$ scales as a power law with exponent $=3/2$. Overlap of two measures demonstrates that nodes typically fail once during each cascade. Network size is $N=5000$, the degree distribution scale-free exponent is $\gamma=3.00$ and the BTW dissipation parameter $f=0.01$. }
\label{fig:Prob_size_area}
\end{figure}

The BTW sandpile model is a prototypical, idealized model of cascading dynamics caused by load shedding on a network \cite{Bak1987,Bak1988}. Throughout this paper, the following formulation of the dynamics is used. Consider a network of $N$ nodes, where each node has some capacity to hold grains of sand, and each grain corresponds to a unit of load. The topology of the network is fixed, while the amount of sand on individual nodes changes in time. The \emph{capacity} of a node is the maximal amount of sand that it can hold. A natural choice is for the node to topple when the amount of sand first equals its degree, $k$ \cite{Noel2014}, as the node can then shed one grain of sand to each neighbor. We therefore set the capacity of each node to $k-1$. Hence, a $(k-1)$-sand node of degree $k$ is \emph{at capacity}, meaning that it holds as much sand (load) as it can withstand. Adding a grain to such a node brings it \emph{over capacity}, and it therefore topples.

The dynamics of the sandpile model consists of a sequence of \emph{cascades} on this network, defined as follows. At each discrete time step, a grain of sand is dropped on a node chosen uniformly at random. If this addition does not bring the initial node over capacity, then that cascade is finished. However, if the node is over capacity, then it topples and sheds one grain to each of its neighbors. Any node that then exceeds its capacity topples in the same way, shedding to its neighbors who may in-turn topple, which continues until all nodes are below or at their capacity (i.e., equilibrium is restored). In order to prevent the system from becoming saturated with sand, a dissipation mechanism is required: whenever a grain of sand is shed from one node to another, it dissipates (is removed) with a small probability $f$. In this paper $f=0.01$, unless stated otherwise. This dissipation rate is chosen so that the largest cascades topple almost the entire network.

The \emph{size} of a cascade is the total number of toppling events, while the \emph{area} of the cascade is the total number of nodes that ever topple. In scale-free networks, we find that these two quantities are essentially the same \cite{Goh2003} (see Fig.~\ref{fig:Prob_size_area}), which is in contrast to the situation for regular random graphs \cite{Noel2013,Noel2014}. As our interest is in scale-free networks, we thus focus on the cascade size in the rest of the paper.

The mean-field solution to the BTW model is characterized by a distribution of cascade sizes, $P(s)$, that exhibits a power law with exponent $-3/2$ \cite{Alstrom1988}. The same scaling behavior is observed for a wide range of networks, from random regular networks and other networks with narrow degree distributions \cite{Bonabeau1995}, to even classes of broad-scale networks \cite{Goh2003}. For scale-free random networks, the distribution deviates from the mean-field result only if the degree distribution is sufficiently heavy tailed. If the exponent of the degree distribution is $2 < \gamma \le 3$, one observes cascade size distribution of exponent $\gamma/(\gamma-1)$ \cite{Goh2003}. Otherwise the mean-field value is observed \cite{Goh2003}. Degree heterogeneity therefore creates either the -3/2 exponent or heavier-tailed cascades, and is not \textit{a priori} a mechanism to reduce the probability of large cascades.

\section{Individual networks}

\begin{figure}[t]
\includegraphics{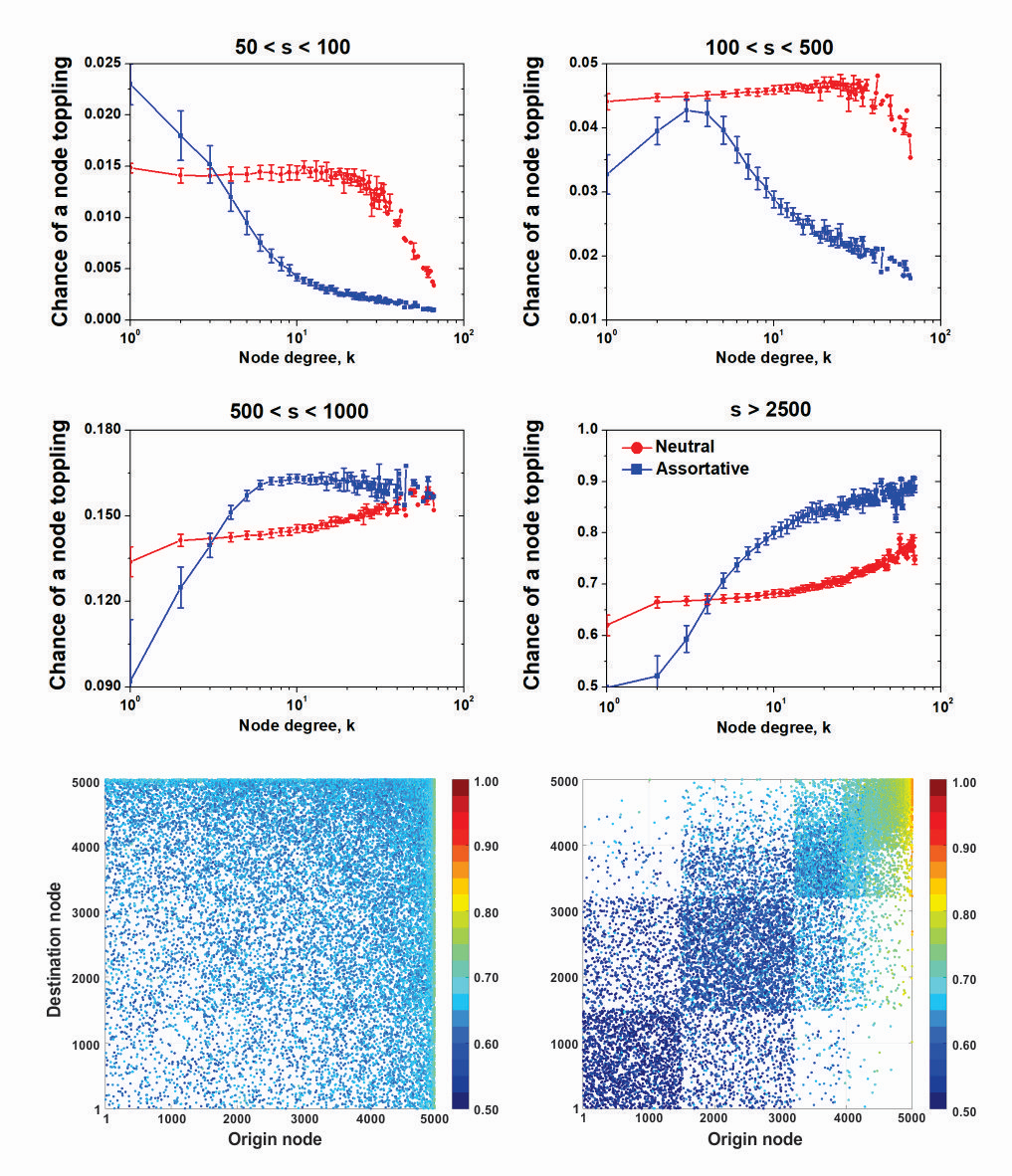}
\caption{Chance of toppling in a cascade of size $s$. Top 4 panels: In a neutral network (red circles), nodes of various degree are approximately equally likely to participate in cascades of different sizes. In an assortative network (blue squares) high degree nodes topple frequently in large cascades and infrequently in small ones, while the reverse is true for low degree nodes. Bottom 2 panels: Adjacency matrices for neutral scale-free network (left) and assortative network (right), where nodes in each figure are sorted from low degree (bottom left corner) to high degree (top right corner). Colors indicate the probability that, in a large cascade, a given link participates in sand redistribution. These panels further demonstrate that, in assortative networks, links connected to high degree nodes are more likely to participate in large cascades. }
\label{fig:Single_topple}
\end{figure}

Individual scale-free networks are generated using a modified version of the configuration model \cite{Satorras2005}. Our results in the main text are for a degree distribution $P(k)\sim k^{-\gamma}$, where $\gamma = 3.00$, the mean degree $\langle k\rangle=4$ and the minimum degree is $k_{min}=2$. We find, however, that the main results appear to not strongly depend on the value of $\gamma$, as seen in the Appendix (cf.~Fig.~\ref{fig:alpha_2x5}). Next, we adopt the rewiring algorithm of \cite{Sokolov2004} to obtain networks with positive correlations between degrees of individual nodes and that of their neighbors. This procedure leads to a modular structure where nodes of similar degree are more likely to be connected with each other. Although the assortativity of the original neutral scale-free networks is low ($a=0.020 \pm 0.005$) \cite{Newman2002}, the rewiring procedure increases assortativity to $a=0.25 \pm 0.02$, which corresponds to significant degree-degree correlations. The assortativity of each network is demonstrated by the block structure of the adjacency matrix showed on Fig.~\ref{fig:Single_topple}. Although our algorithm can produce assortativity as high as $a=0.45 \pm 0.05$, we choose a more moderate assortativity to reduce degree-induced modularity, and to better match the assortativity of empirical networks \cite{Braun2012}.

In Fig.~\ref{fig:Single_topple} we demonstrate how the probability for nodes to topple changes with cascade size for both neutral and assortative networks. In assortative scale-free networks, low degree nodes topple more often in small cascades than in large ones. The opposite is true for high degree nodes. We therefore infer that high degree nodes must topple if a large cascade is to occur. In comparison the likelihood for nodes to topple in neutral networks is much less dependent on degree.

\section{Sand Distribution Assumption} \label{sec:one_over_k}

In the previous section, we notice the relationship between cascade size and node degree is different for neutral and assortative scale-free networks. In particular the almost constant probability for nodes to topple in the neutral topology comes as a surprise, because it contradicts the existing assumption that the probability that a node topples is proportional to the inverse of a node's degree, which is commonly referred to as the $1/k$ ansatz. This ansatz is often a fundamental assumption when determining how the dynamics approach a critical branching process \cite{Bonabeau1995,Goh2003,Noel2014}. Because we observe that high-degree nodes topple with a relatively high probability that strongly diverges from $1/k$, we are motivated to investigate the validity of the $1/k$ ansatz. In \cite{Noel2014} the authors do show that the $1/k$ assumption is not strictly valid and that higher degree nodes are more likely to be at capacity. Yet, they attribute the observed power law distribution of failures sizes to be due to universality. We shown below that analyzing the out of equilibrium distribution of sand on nodes for the BTW model provides the resolution for how a critical branching process can arise from a system seemingly poised for a super-critical branching process.

\subsection{Equilibrium configuration of the sandpile model}

\begin{figure}
\includegraphics[width=8.6cm]{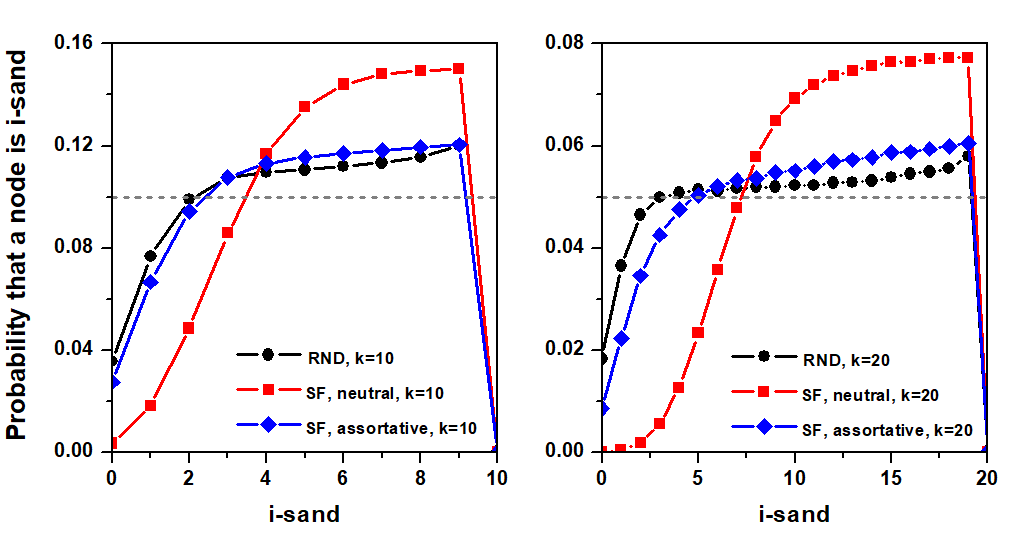}
\caption{A corollary of the ``$1/k$ ansatz'' is that the probability for a degree-$k$ node to have $i$ grains of sand is constant and equal to $1/k$, as denoted by the dashed line. However, for $k$-regular networks, as well as neutral and assortative scale-free networks, behavior deviates strongly from this corollary. Degree$-10$ nodes (left panel) and degree$-20$ nodes (right panel) are more likely to have near-critical amounts of sand, and less likely to have low amounts of sand. Furthermore, nodes in assortative scale-free networks behave similarly to nodes in $k$-regular graphs of the same degree, while nodes in neutral scale-free networks show the strongest deviation from the $1/k$ ansatz, with loads skewed strongly towards critical capacity. Network size is $N=5000$ and the degree distribution scale-free exponent is $\gamma=3.00$.}
\label{fig:iSAND_rand_sf}
\end{figure}

We begin by considering the distribution of sand on the network in equilibrium (just before an additional grain of sand is dropped). A corollary of the $1/k$ ansatz is that the probability for a degree-$k$ node to have $i$ grains of sand is constant and equal to $1/k$, such that there is no typical amount of sand in any inactive node \cite{Bonabeau1995,Goh2003,Noel2014}. In Fig.~\ref{fig:iSAND_rand_sf}, we notice that the distribution of sand differs significantly from analytic assumptions. Although past works \cite{Bonabeau1995,Goh2003,Noel2014} report that the $1/k$ corollary approximately holds, our observations demonstrate a more complex picture. As shown in Fig.~\ref{fig:iSAND_rand_sf}, the probability that a degree-$k$ node has $i$-sand is not $1/k$, but is strongly skewed towards larger values of $i$. This observation holds for various node degrees, network topologies, network sizes, and various values of the dissipation rate, $f$, as discussed in detail in the Appendix.

Furthermore, we observe a strong effect of degree correlations: degree$-k$ nodes in neutral and assortative scale-free networks are characterized by different sand distributions. The departure from $1/k$ distribution is particularly pronounced for the neutral topology, where probability for a node to be close to capacity is nearly twice what we would expect theoretically.

Additionally we notice strong similarities between $i$-sand distributions for $k$-regular and assortative networks. This property, combined with the modular nature of assortative network, suggests that modules of similar degree dynamically behave like regular graphs with the same degree. One could interpret the BTW dynamics on assortative scale-free networks as one on a set of coupled regular graphs of increasing degree. We will demonstrate later that this property has significant impact on the occurrence of catastrophic failures in interconnected heterogeneous networks.

In summary, the probability that a node topples (Fig.~\ref{fig:Single_topple}), and the probability that a node has $i$ grains of sand (Fig.~\ref{fig:iSAND_rand_sf}) both show strong deviations from theoretical assumptions previously relied upon to explain the critical dynamics of the BTW model. Because empirically we observe that the probability of a node toppling is greater than $1/k$, it would naively imply that the BTW model always produces a super-critical branching process. Why, then, does all past research observe a power-law tail in the distribution of cascade sizes, a signature of a critical process? Figure~\ref{fig:Prob_size_area}, for example, shows that the cascade size and area distribution for scale-free networks broadly follows a power-law distribution over several orders of magnitude. Larger networks produce even stronger agreement with the theoretical power-law distribution.

\subsection{Dynamics out of equilibrium}

\begin{figure}[t]
\includegraphics[width=8.6cm]{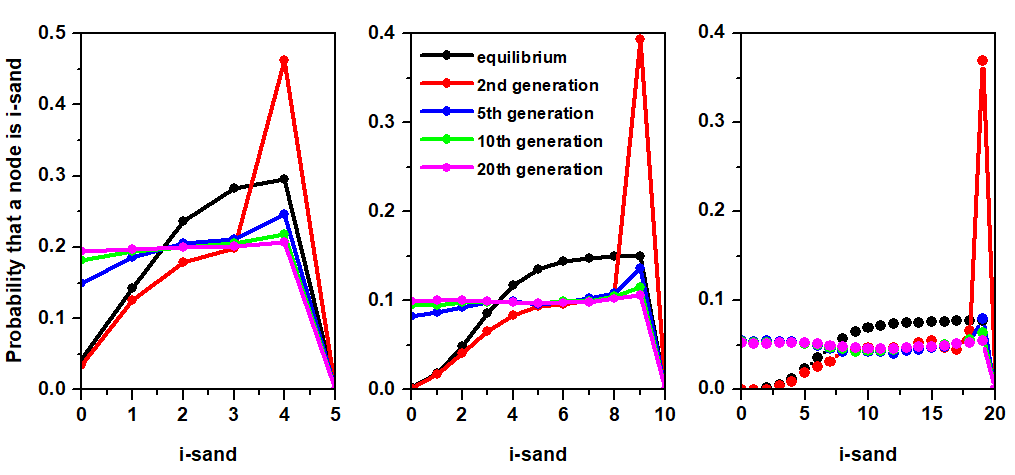}
\caption{
Distribution of sand in and out of equilibrium. Left panel: degree$-5$, middle panel: degree$-10$, and right panel: degree$-20$ nodes. Black lines are the equilibrium distribution, just before any sand is dropped. Red, blue, cyan and magenta lines correspond to the sand distribution among nodes that receive grains of sand over the course of large cascades, where $s>N/2$ nodes topple. As the cascade progresses (from early generations to later ones), the sand distribution approaches the $1/k$ corollary, and sand topples with probability $1/k$ in agreement with the ansatz. In the figures, the networks are neutral scale-free with $\gamma=3.00$ and size $N=5000$. Other topologies show similar behavior.}
\label{fig:iSAND_sf}
\end{figure}

To understand why the BTW model creates power-law distributions, we study out-of-equilibrium behavior of the model's dynamics. Namely, we investigate the probability of a degree-$k$ node having $i$-sand as a large cascade progresses. We consider a 
cascade evolution scheme that proceeds according to generations, in parallel with the nomenclature of the branching processes. The node toppling as a result of the initial random deposition of a grain of sand is called a \textit{root} and forms the first generation of the cascade. Each successive generation is formed by the nodes that received sand from the previous generation's nodes that have toppled. Figure~\ref{fig:iSAND_sf} demonstrates our results. We find that, regardless of node degree, the sand distribution on nodes in the $n$\textsuperscript{th} generation is a better and better approximation of the analytic assumption of $1/k$.

Initial generations strongly disagree with theory. For example, there is a clear peak in the distribution at second generation because, in order for a cascade to be large, a high number of neighbors must topple. These initial generations, however, consist of very few nodes. The second generation consists of at most $k$ nodes, and the third generation has less than $k\times \langle k\rangle_{nn}$ nodes on average, where $\langle k\rangle_{nn}$ is the mean degree of the nearest neighbors. Each generation $g$ is bounded by $k\times (\langle k\rangle_{nn})^{g-2}$ thus the bulk of the nodes in large cascades are those in the later generations that happen to closely approximate the $1/k$ assumption.

In summary, the BTW dynamics are characterized by the equilibrium sand distribution differing from $1/k$ ansatz, but, over the course of a cascade, the sand distribution among nodes that receive sand evolves to the theoretically expected distribution. There are two properties of large cascades that allow the observed BTW dynamics to more closely approximate the prediction of the theory. First, the out-of-equilibrium statistics are only based on nodes that receive sand, implicitly introducing a biased sub-sampling of sand distribution on the nodes. This is why the equilibrium distribution disagrees with the $1/k$ assumption. Second, we observe the nodes that toppled in earlier generations receive sand in future generations, but never enough sand to topple a second time. For example, a node that topples during a cascade will send sand back to the parent node that toppled it. Additionally, due to non tree-like structure of the network, a non-negligible fraction of nodes receive two and more grains of sand per generation. Thus loops present in the network affect the sand distribution on the network, and are responsible for skewed equilibrium $i$-sand distribution.

\section{Interconnected scale-free networks}  \label{sec:sandpile_conn}

In this section, we study how network topology affects the probability of large cascades with various intra- and interlayer connectivity statistics. We focus on the exceptionally large cascades in the BTW sandpile model due to the disproportionate cost associated with such large extreme events, when compared to the cost of small events, occurring in real interconnected systems. We generate two scale-free networks, denoted here network (or layer) A and network (or layer) B, each with $N$ nodes ($2\times N$ nodes total). We connect nodes within layers either at random or assortatively, and connect nodes between layers either at random or in a hub-to-hub fashion. Random inter-layer connectivity is created by connecting $p\times N$ random pairs of nodes together between layers. Hub-to-hub inter-layer connectivity is created with the following algorithm:
\begin{enumerate}
\item Sort nodes in each network from highest-degree to lowest-degree
\item Connect $p\times N$ nodes in each network together starting with the highest-degree node and working down.
\end{enumerate}
The parameter $p$, which varies between 0 and 1, dictates the strength of coupling between individual networks.

\subsection{Assortative scale-free networks}

\begin{figure}[t]
\includegraphics[width=8.6cm]{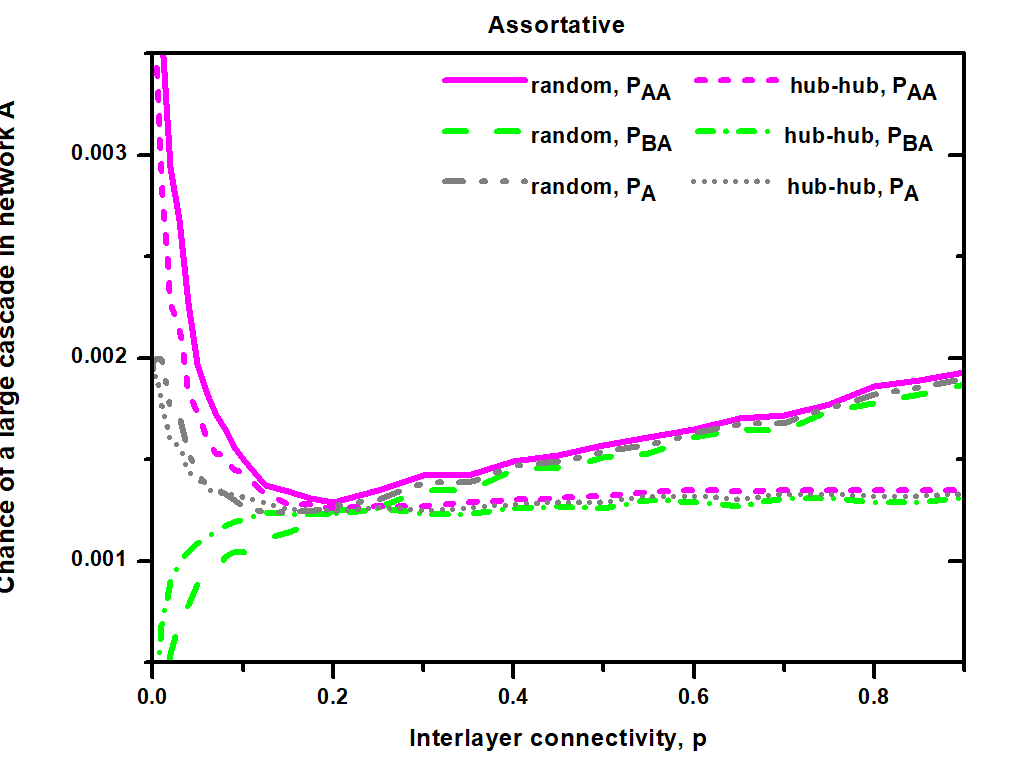}
\caption{The probability of a large cascade for two coupled assortative scale-free networks versus the connectivity probability $p$. Inter-layer connectivity has a strong impact on the BTW dynamics, with the hub-to-hub coupling resulting in a constant chance of cascading failures, while random coupling results in an increasing occurrence of large events. Each network has $N=5000$ and the dissipative parameter $f=0.01$.
}
\label{fig:Prob_ALL_ass}
\end{figure}

We first focus our attention on the effect that the presence of other layers has on an individual network layer (without loss of generality, we can choose layer A). For each cascade, we record the number of nodes that toppled during the process, $s_A$ and $s_B$, separately for both layers. We then determine the probability $s_A>N/2$. We differentiate between the chance of large cascades occurring locally in layer A, denoted $P_{AA}(s_A>N/2)$, and that of cascades originating in layer B \emph{(inflicted cascades)} and traversing into layer A, denoted $P_{BA}(s_A>N/2)$. This allows us to understand how local cascades spread within and across networks.

In Fig.~\ref{fig:Prob_ALL_ass}, we find that the probability of large cascades in coupled assortative networks depends strongly on the mode of interlayer coupling. With hub-to-hub coupling, we notice that $P_{AA}(s_A>N/2)$ initially drops dramatically with increasing interlayer coupling and, furthermore this probability barely changes for $p>0.2$. In comparison, $P_{BA}(s_A>N/2)$ increases initially but also reaches a constant value for $p>0.2$. The overall probability of large cascades stays thus constant for $p>0.2$. With random inter-layer coupling, however, moderate to high inter-layer connectivity significantly increases the likelihood of large cascades that originate in both layer A or B. This latter behavior resembles one observed in random regular networks examined by Brummitt et al.~\cite{Brummitt2012}, suggesting a similar mechanism.

\begin{figure}[t]
\includegraphics[width=8.6cm]{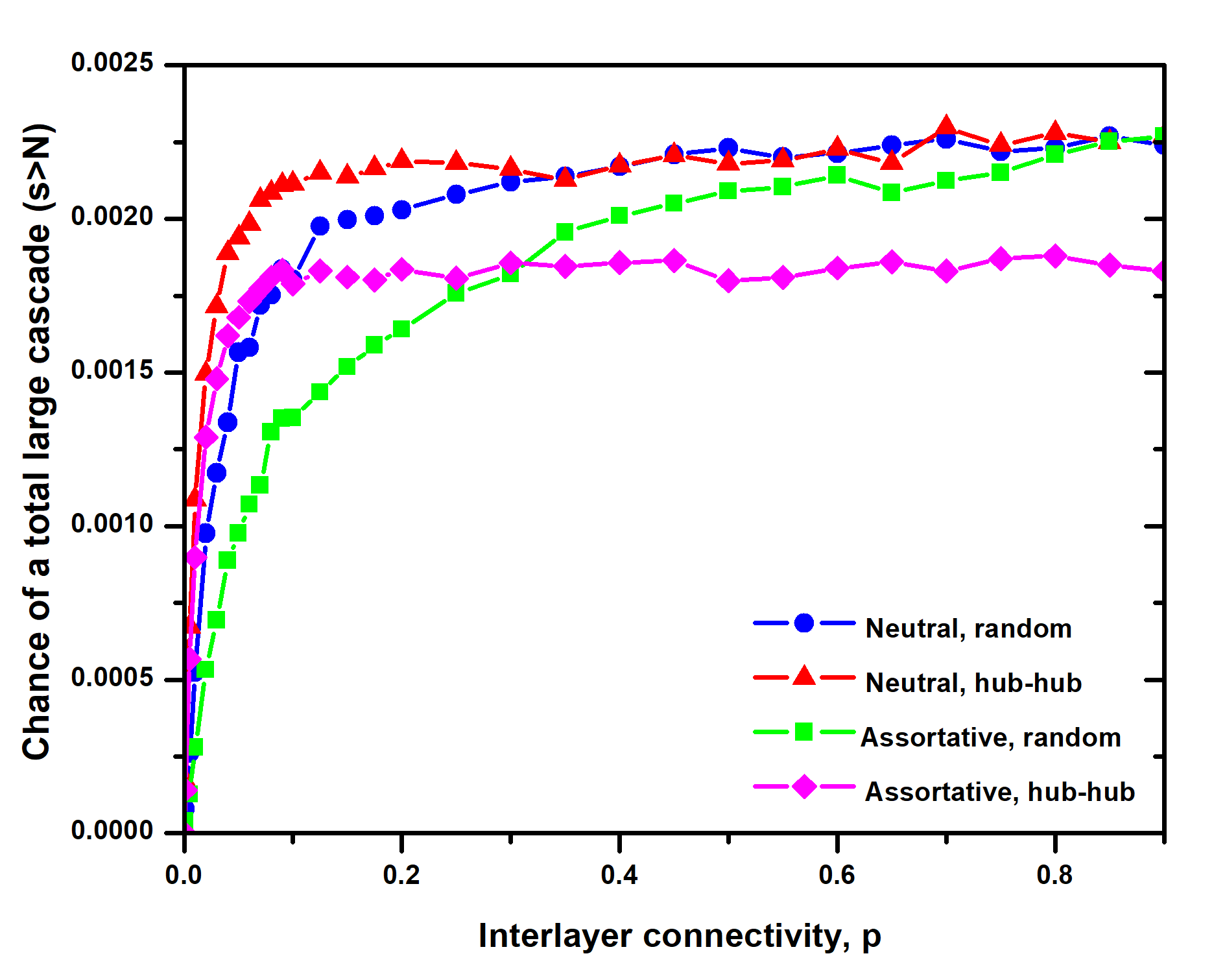}
\caption{The probability of a cascade greater than $N=5000$ versus inter-layer connectivity $p$. Shown are neutral networks with random and hub-to-hub coupling as well as assortative networks with random and hub-to-hub coupling.}
\label{fig:Prob_cascade_total}
\end{figure}

The decrease in the likelihood of large cascades for hub-to-hub coupled assortative networks lies in their highly modular structure (see connectivity of nodes in Fig.~\ref{fig:Single_topple}). Hub-to-hub connectivity extends the modular structure of individual layers to the entire dual-layer system, preserving linkage of nodes of similar degree. Furthermore, in an individual assortative layer, the occurrence of a large cascade is conditioned on toppling of several high degree nodes (see Fig.~\ref{fig:Single_topple}), a rare event. It is quadratically less probable that such condition will be met in a double layer system. Coupling layers of assortative scale-free networks in a hub-to-hub fashion therefore decreases the likelihood of large cascades by absorbing excess load from a layer.

In contrast, by coupling the layers of assortative networks randomly, there is a greater amount of connectivity between degree-$k$ modules, and the connectivity gives rise to a more homogeneous network, reminiscent of coupled random regular networks studied by \cite{Brummitt2012}. Following Brummitt \emph{et al.}~\cite{Brummitt2012}, we believe that an increase in the probability of large cascades for large $p$ is caused by diverted loads that return to the network. Random wiring allows for high-degree nodes in one layer to connect to low-degree nodes in another, which allows for cascades to more easily cross layers because low-degree nodes topple more often. Because low-degree nodes shed their sand more often, there is also a greater likelihood for high-degree nodes to topple, thereby triggering a large cascade.

We can extend our results to large cascades affecting both layers of the network as well. We show in Fig.~\ref{fig:Prob_cascade_total} that for $p>0.3$, the regime where large cascades are suppressed, assortative, hub-to-hub coupled networks have the smallest probability of large cross-network cascades where $s>N$. Predictably, however, greater connectivity increases the probability of large cascades overall. In real systems, however, it may be important to connect all regions together, therefore the assortative hub-to-hub topology produces the best trade-off of inter-connectivity without as large a probability of large cascades.

\subsection{Neutral scale-free networks}

\begin{figure}
\includegraphics[width=8.6cm]{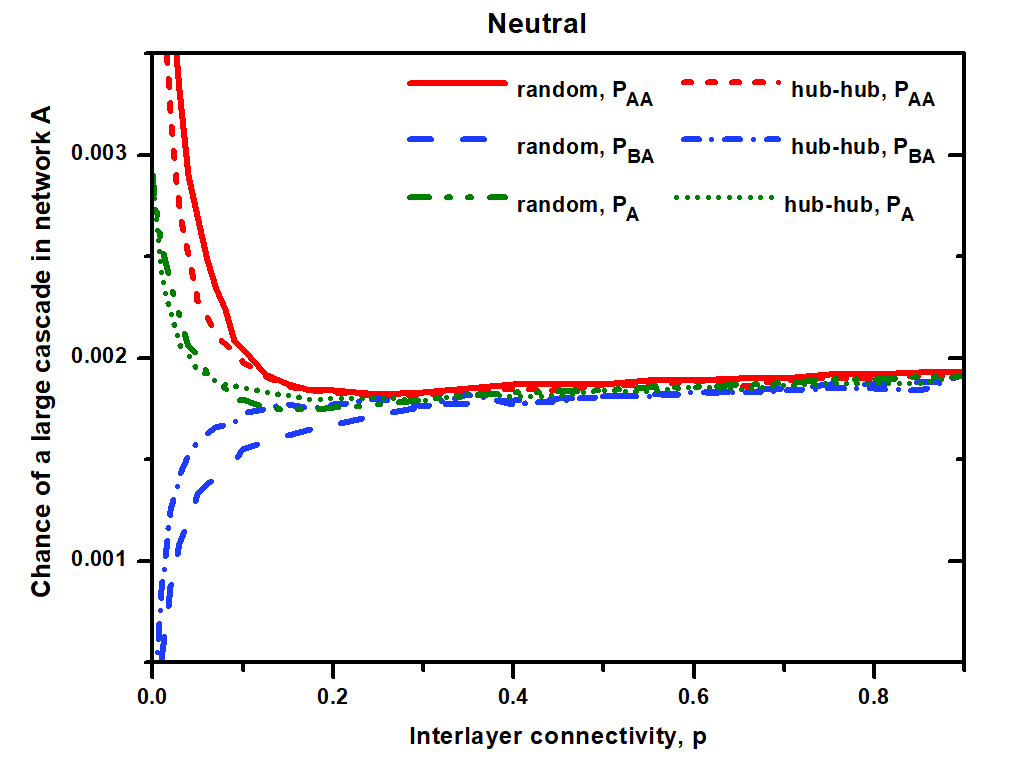}
\caption{The probability of a large cascade occurring in a system of two coupled neutral scale-free networks versus the connectivity probability, $p$. This probability is virtually indistinguishable between hub-to-hub and random interlayer connectivity. As the connectivity probability $p$ increases, cascading failures reach a constant. In each network, $N=5000$ and the dissipative parameter $f=0.01$. }
\label{fig:Prob_ALL_neutral}
\end{figure}

In this section, we focus on coupled random scale-free networks. This represents networks with the least intra-layer structure. As illustrated in Fig.~\ref{fig:Prob_ALL_neutral}, the probability for large cascades that originate in network A drops substantially with $p$, while the probability of large inflicted cascades rises. The overall probability that any cascade occurs in network A, $P_A$, is reduced with introduction of interlayer coupling, although for $p>0.3$, the probability is roughly constant, similar to assortative scale-free networks coupled hub-to-hub. The mechanism responsible for this behavior, however, differs fundamentally from that in the case of assortative networks.

Namely, the lack of degree correlations in neutral scale-free network causes the toppling rate to be almost independent of node degree (Fig.~\ref{fig:Single_topple}), regardless of cascade size. Thus, interlayer coupling, whether random or hub-to-hub, has approximately the same effect on the dynamics. With low coupling, network A benefits from shedding load to network B, but once there is moderate coupling, the two networks act as a single random network, implying the probability of large cascades does not vary significantly for $p>0.2$.

Finally, we notice that random networks produce higher probabilities of large cross-network cascades compared to assortative networks, as seen in Fig.~\ref{fig:Prob_cascade_total}. This further demonstrates the rationale for assortative networks, and suggests that real systems undergoing failure cascades may be evolutionarily disinclined to be random.

\section{Discussion}\label{sec:discussion}

We set out to better understand the dynamics of the BTW sandpile model, a prototypical SOC model \cite{Bak1987,Bak1988}. In doing so, we first noticed an under-appreciated aspect of the model: the node sand distribution differs markedly from the theoretical assumption. The distribution would seemingly imply that the dynamics are super-critical in equilibrium, but the non-equilibrium statistics demonstrate that the sand redistributes to create critical dynamics.

Although the BTW model is simplistic, it creates important insights into how the spread of cascading failures is affected by underlying network topology. We demonstrate that the robustness of interconnected systems is a function of correlations between intra- and interlayer interactions. This is similar to earlier results of Reis \textit{et al}. \cite{Reis2014}, based on studies of bond percolation-like process on interconnected scale-free networks. Intriguingly, the network topology that we found most effective in reducing large cascades---assortative scale-free networks with hub-to-hub inter-connectivity---is found in functional brain networks \cite{Eguiluz2005,Bullmore2009,Bialonski2013,Geier2015,Heuvel2011,Reis2014}. Because neuronal avalanches appear to self-organize to a critical state \cite{Levina2007,Goh2003,Beggs2003,Chialvo2010}, the human brain can be modeled in a simplified manner via the BTW model. Taking these results together, we would predict that the brain is constructed so as to prevent large cascades. If we were to interpret large cascades as seizures, for example, this would make the surprising suggestion that a healthy brain naturally reduces the likelihood of seizures, and a reduction in assortativity, or hub-to-hub inter-connectivity would make seizures more likely. This agrees with recent findings that particular abnormal functional brain networks, such as those observed in schizophrenia \cite{Heuvel2013,Collin2014}, increase the likelihood of seizures \cite{Hyde1997,Clancy2014}. Furthermore, in agreement with our model (Fig.~\ref{fig:Single_topple}), rich club nodes (hub nodes in assortative networks) are strongly associated with generation seizures \cite{Lopes2017}. Overall, our results show that despite the BTW model's simplicity, it can qualitatively approximate the occurrence of brain seizures. Moreover, it can provide insight into the evolutionary motivation of functional brain network topologies.

Our work suggests several avenues of future research. We find that high-degree nodes in assortative networks are likely to topple in large cascades, therefore one could design protocols controlling the amount of load on those nodes or devise quarantine scenarios in order to limit the spread of catastrophic failures. Moreover, a fruitful avenue of research would be predicting large cascades when a cascade is beginning, such as detecting whether hub nodes shed their load. In addition, we only explored this behavior for the BTW model. It is an open question to understand if these same three features of heterogeneous degree distributions, internal network assortativity and interlayer degree correlations also suppress large-scale failure for other types of cascade models, such as threshold models.

\section{Acknowledgements}
We gratefully acknowledge support from the US Army Research Office MURI award W911NF-13-1-0340 and Cooperative Agreement W911NF-09-2-0053 (Network Science CTA); The Minerva Initiative award W911NF-15-1-0502; and DARPA award W911NF-17-1-0077.

\section{Appendix} \label{sec:supp_mat}

In this section, we discuss several details of the model that, in the interest of space, we leave out of the main text.

\begin{figure}
\includegraphics[width=8.6cm]{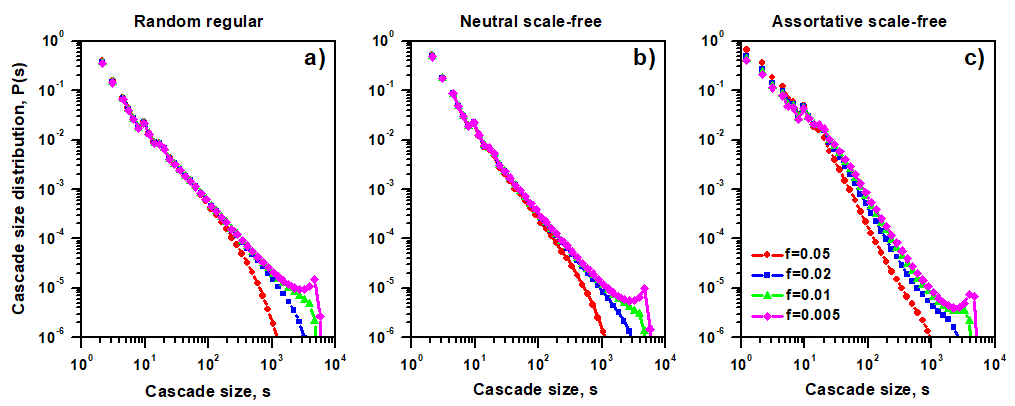}
\caption{ Choice of the dissipation rate $f$ adopted in the sandpile dynamics affects the tail of the distribution of cascade sizes, $P(s)$. }
\label{fig:Ps_DISS}
\end{figure}

\begin{figure}
\includegraphics[width=8.6cm]{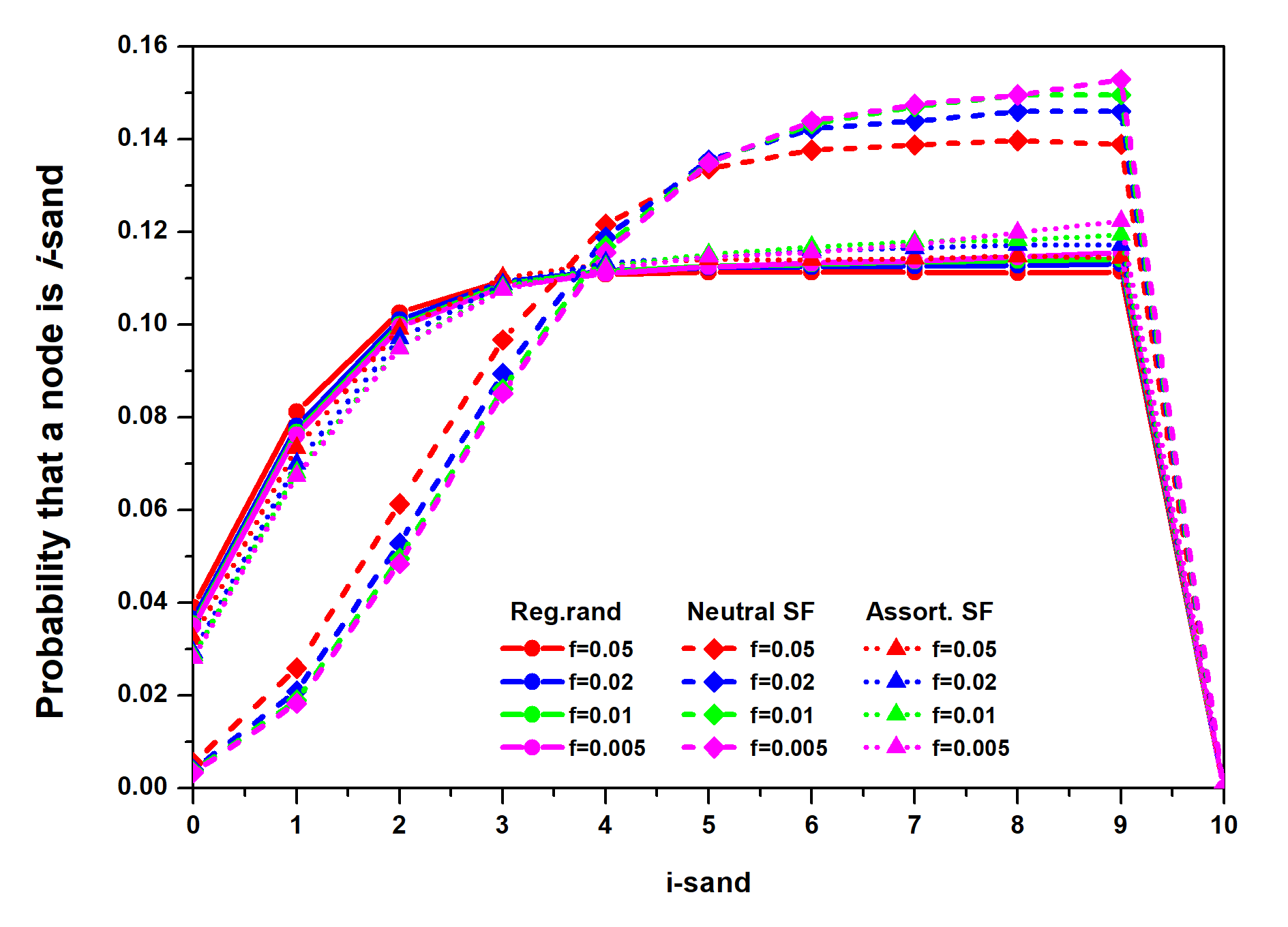}
\caption{Probability that a degree-$k$ node holds $i$ grains of sand stays constant despite significant variation in the dissipation rate $f$ of the sandpile process. Behavior of a random $10$-regular network is compared with $k=10$ nodes of a neutral and assortative scale-free network, respectively.  }
\label{fig:iSAND_DISS}
\end{figure}

\subsection{Dissipation rate and system size.}

After observing the sand distribution shown on Fig.~\ref{fig:iSAND_rand_sf}, which differs from theoretical assumption of critical branching process, one might suspect that presented results are side effects of one of the model's two parameters: the dissipation rate, $f$, or the system size, $N$. However, varying those parameters does not appear to better approximate the $1/k$ corollary.

First we consider the effect of different values of the dissipation rate $f$ on the sandpile dynamics. Since this constant regulates total amount of sand on the network, higher values correspond to increased sand removal, while lower values lead to the accumulation of load in the system. This behavior is reflected in the statistics of observed cascades, illustrated in Fig.~\ref{fig:Ps_DISS}. The former condition of lowering load results in decreased probability of large cascades, denoted by a truncation of the tail of the cascade size distribution. On the other hand excessive sand accumulation gives rise to more frequent large cascades, as shown by a visible peak in the $P(s)$ function for $s\sim O(N)$.

However, even though varying $f$ significantly changes the cascade size distribution, it has little effect on the sand distribution on individual nodes (Fig.~\ref{fig:iSAND_DISS}). As lower value of $f$ leads to sand accumulation, we observe slight increase in probability of node being at capacity, but the effect is very subtle, especially when contrasted with the impact that change in $f$ has on macroscopic observables, such as $P(s)$. As noted earlier, the distribution of sand on nodes of assortative scale-free network overlaps with the distribution of sand on random regular network of the same degree, independent of the selected value of $f$.

\begin{figure}
\includegraphics[width=8.6cm]{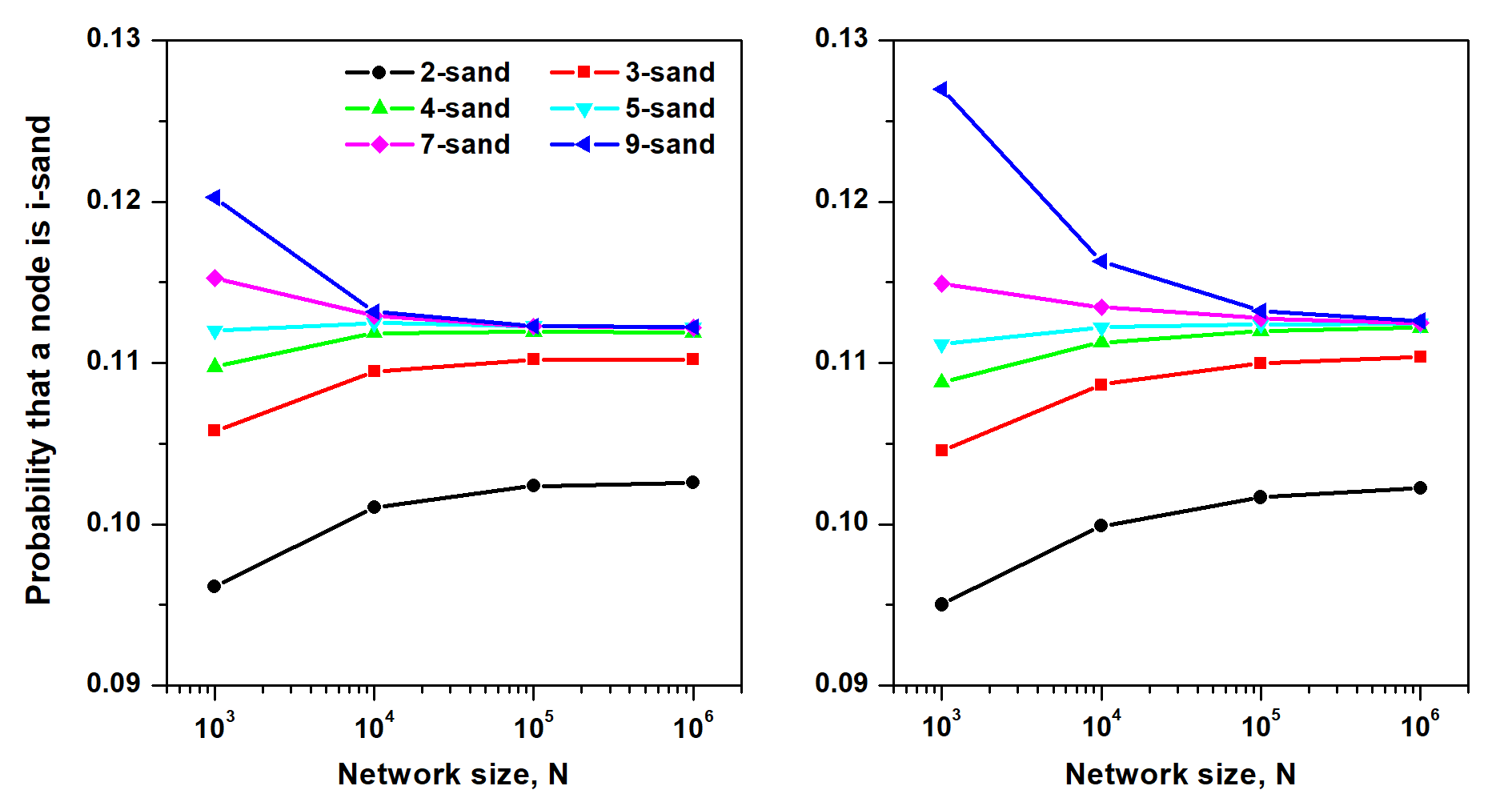}
\caption{Probability that a degree-$k$ node holds $i$ grains of sand saturates as a network size $N$ increases. Behavior of a sandpile process on a random $10$-regular network with dissipation rate $f=0.01$ and $f=0.001$ is shown on the left and right panel, respectively.}
\label{fig:iSAND_N}
\end{figure}

Finally, we consider the impact of finite system size, $N$. In Fig.~\ref{fig:iSAND_N} we show, that despite increasing the system size by three orders of magnitude, the disparities in sand distribution are preserved.

\subsection{Exponent of the scale-free network degree distribution.}

\begin{figure}[t]
\includegraphics[width=8.6cm]{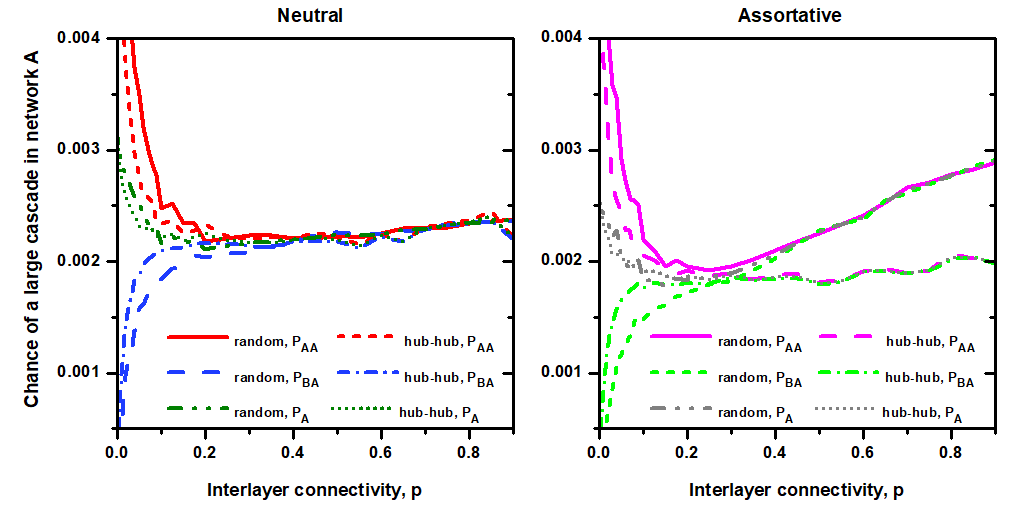}
\caption{Chance of large cascades occurring in a system of randomly connected scale-free networks, where scaling exponent of the individual layer degree distribution is $\gamma=2.50$. Despite differences in absolute probability values, qualitative behavior remains the same as for $\gamma=3.00$ (compare to Fig.\ref{fig:Prob_ALL_neutral} and Fig.\ref{fig:Prob_ALL_ass}). }
\label{fig:alpha_2x5}
\end{figure}

 All results reported in the main text refer to scale-free networks with a degree distribution $P(k)\sim k^{-3}$, therefore we want to test if heavier-tail distributions affect significantly those observations. However, we find that, e.g., for $P(k)\sim k^{-2.5}$, our results appear similar to those discussed in the main text. Despite differences in absolute values of probabilities of large cascades, qualitative behavior for all considered intra- and interlayer coupling modes is the same, as shown in Fig.~\ref{fig:alpha_2x5}.

\bibliographystyle{apsrev4-1}
\bibliography{SFCascades}

\end{document}